\documentclass[journal]{IEEEtran}
\usepackage[compress]{cite}
\usepackage{graphicx}
\usepackage{amsthm, amsmath}
\usepackage{url}
\usepackage{subcaption}
\usepackage{algorithm}
\usepackage{algpseudocode}
\usepackage{algcompatible}
\usepackage[super]{nth}
\usepackage[english]{babel}
\usepackage{multirow}
\usepackage{makecell}
\usepackage{soul}
\usepackage{color}
\renewcommand{\baselinestretch}{.97}

\begin{document}

\title{Opportunistic Algorithm for Data Ferrying in Smart Communities with Limited Communications Infrastructure}
\title{Opportunistic Data Ferrying in Areas with Limited Information and Communications Infrastructure}

\author{Ihab Mohammed~\IEEEmembership{Student Member,~IEEE}, Shadha Tabatabai~\IEEEmembership{Student Member,~IEEE}, Ala Al-Fuqaha~\IEEEmembership{Senior Member,~IEEE}, and Junaid Qadir~\IEEEmembership{Senior Member,~IEEE}}

\maketitle

\begin{abstract}


Interest in smart cities is rapidly rising due to the global rise in urbanization and the wide-scale instrumentation of modern cities. Due to the considerable infrastructural cost of setting up smart cities and smart communities, researchers are exploring the use of existing vehicles on the roads as ``message ferries'' for the transport data for smart community applications to avoid the cost of installing new communication infrastructure. In this paper, we propose an opportunistic data ferry selection algorithm that strives to select vehicles that can minimize the overall delay for data delivery from a source to a given destination. Our proposed opportunistic algorithm utilizes an ensemble of online hiring algorithms, which are run together in passive mode, to select the online hiring algorithm that has performed the best in recent history. The proposed ensemble-based algorithm is evaluated empirically using real-world traces from taxies plying routes in Shanghai, China, and its performance is compared against a baseline of four state-of-the-art online hiring algorithms. A number of experiments are conducted and our results indicate that the proposed algorithm can reduce the overall delay compared to the baseline by an impressive 13\% to 258\%.

\end{abstract}

\IEEEpeerreviewmaketitle

\section{Introduction}

It is estimated that about 60\% of the world's population will live in cities by 2030 \cite{undata2016}. Additionally, by the year 2020, around 20.4 billion devices are expected to be connected to the Internet \cite{Tabatabai_2017}. To cope with the trend of people moving to urban centers and to provide high-quality services to their residents, municipalities are increasingly turning to Information and Communications Technologies (ICT)---such as cloud computing, Internet of Things (IoT), Wireless Sensor Networks (WSNs), and Cyber-Physical Systems (CPS)  \cite{net_Comm_2017}---for the deployment of smart community applications that provide value-added services in diverse fields such as healthcare, transportation, entertainment, and governance~\cite{ala_smartCity2017}.

But such smart community deployments are often prohibitively expensive, especially for smaller communities where the deployment of smart community applications is hindered by the unavailability of appropriate communication infrastructure. One way to address this concern is to exploit existing infrastructure in innovative ways. In particular, modern vehicles that abundantly ply the roads of urban cities can be exploited to obviate the need for an expensive communications infrastructure. In recent times, with increased interest in vehicular ad-hoc networks (VANETs) and self-driving cars, vehicles are increasingly becoming more sophisticated and it is expected that by the year 2020, 90\% of vehicles will be equipped with a hardware-based On-Board Unit (OBU) \cite{OBU_2017} that has processing and communications capabilities. Therefore developing an approach for opportunistically accessing these smart vehicles in an efficient delay-tolerant manner becomes a promising approach towards the deployment of cost-effective and efficient smart community applications with limited ICT infrastructure overhead. For example, rural areas that lack the funds to deploy ICT infrastructure can benefit from our proposed approach.

One approach of delivering data in sparse networks is Message Ferrying (MF). In this approach, devices are classified as \textit{message ferries} (or \textit{ferries}) or \textit{regular nodes}. Ferries are mobile devices that move around to collect messages from regular nodes to deliver them to their destination~\cite{zhao2004message}. In this paper, we use this approach and utilize vehicles as ferries to transfer data collected from the smart devices (i.e., regular nodes) to the Smart Community Management Center (SCMC). One example of the SCMC is the Traffic Management Center (TMC), which is used for managing traffic in support of intelligent transportation applications.

In this paper, we envision a smart community application architecture where the service area is divided into blocks, each having at least a single Local Community Broker (LCB), where an LCB is a cloudlet that is deployed in the block or hosted on a vehicle that has processing and communications capabilities. Each LCB serves as a block manager responsible for selecting vehicles to transfer data collected from IoT devices scattered across the block to the Smart Community Management Center (SCMC). When a vehicle passes through the block, the block manager (i.e., LCB) acquires the estimated delivery delay from the vehicle and decides on whether to utilize it as a ferry to transfer collected data to the SCMC. Consequently, the vehicles themselves are used as ferries to transfer data between the different blocks of the service area. Actually, sparsely deployed LCBs are the only required infrastructure in our proposed architecture and no communications infrastructure (optical, microwave, or 5G base stations) is required to relay the data from the LCBs to the SCMC. 

The main challenge in this architecture lies in the `intelligent' selection of vehicles. If the block manager is not selective, it may end up using vehicles with high delivery delay as any vehicle might be selected even if it has a high delivery delay---delivery delay is the time taken to transfer a data bundle on a vehicles selected to serve as a data ferry from the LCB to the SCMC. On the other hand, if it is very selective and picks vehicles that have a short delivery delay, it would incur high waiting delay as it needs to wait longer to find such vehicles---waiting delay is the time taken to select a vehicle to serve as a data ferry. In fact, this problem is of an online nature because once a vehicle leaves the block, the block manager can no longer utilize the vehicle. Consequently, the block manager may regret its decision when the delivery delay of future vehicles is shorter than that of previous ones.

To solve this problem, we propose an online algorithm that utilizes an ensemble of online hiring algorithms. The proposed algorithm opportunistically selects the best performing algorithm from its ensemble based on the recent observed performance. The average overall delay is the sum of the average waiting delay and the average delivery delay as detailed later in the paper. The proposed algorithm selects one algorithm from its ensemble to be active with the objective of minimizing the average overall delay. All other algorithms in the ensemble are set to be passive (i.e., inactive). This way, the active algorithm alone selects the vehicles to be used as data ferries. Although other algorithms in the ensemble will not be utilized to select the data ferries, their performance is utilized in order to choose the best performing algorithm in the ensemble by choosing the one that demonstrated the lowest average overall delay in the recent history (i.e., greedy approach). Consequently, different hiring algorithms from the ensemble might be utilized over time.

To the best of our knowledge, this is the first research effort that utilizes online hiring algorithms for the selection of data ferries in the specific setting of smart community applications. We perform a thorough evaluation of our work and show that our proposed algorithm performs better than other state-of-the-art online hiring algorithms in a wide variety of settings (including for different traffic volumes).

\section{Related Work}

The literature is rich with research that deals with network issues in smart communities. Jawhar et al. \cite{net_Comm_2017} presented the networking requirements for different smart city applications and additionally presented network architectures for different smart city systems. In \cite{ala_smartCity2017}, the authors discussed the networking and communications challenges encountered in smart cities. Paradells et al. \cite{infrastructureless_2014} state that deploying wireless sensor networks along with the aggregation network in different locations in the smart city is very costly and consequently propose an infrastructure-less approach in which vehicles equipped with sensors is used to collect data. 

Bouroumine et al. \cite{theInfluence_2016} present a system where public and semi-public vehicles are used for transporting data between stations distributed around the city and the main server. Aloqaily et al. \cite{continuousService_2017} introduce the concept of Smart Vehicle as a Service (SVaaS). They predict the future location of the vehicle in order to guarantee a continuous vehicle service in smart cities. In another work \cite{howFeasible_2016}, the authors indicate that cars will be the building blocks for future smart cities due to their mobility, communications, and processing capabilities. They propose Car4ICT, an architecture that uses cars as the main ICT resource in a smart city. The authors in \cite{dataCollection_2017} propose an algorithm for collecting and forwarding data through vehicles in a multi-hop fashion in smart cities. They proposed a ranking system in which vehicles are ranked based on the connection time between the OBU and the RSU. The authors claim that their ranking system results in a better delivery ratio and decrease the number of replicated messages.

In~\cite{DTN_bus_mobility_2016}, authors state that existing network infrastructure in smart cities can not sustain the traffic generated by sensors. To overcome this problem, an investment in telecommunication infrastructure is required. However, authors proposed to exploit buses in a Delay Tolerant Network (DTN) to transfer data in smart cities. In \cite{mobileCloud_2017}, the authors introduce mobile cloud servers by installing servers on vehicles and use them in relief efforts of large-scale disasters to collect and share data. These mobile cloud servers convey data among isolated shelters while traveling and finally returning to the disaster relief headquarters. Vehicles exchange data while waiting in the disaster relief headquarters, which is connected to the Internet.

Bonola et al. \cite{opportunistic_2016} conduct a study on using taxi cabs as oblivious data mules for data collection and delivery in smart cities. They have no guarantee on data communications since they are using taxi cabs without any selection criteria. They use real taxi traces in the city of Rome and divide the city into blocks of size $40 \times 40$ meter$^2$. Depending only on opportunistic connections between vehicles and nodes, the authors claim achieving a coverage of 80\% of the downtown area over a 24 hour period.

The aforementioned papers mostly utilize multiple relays for transferring data between source-destination locations. Furthermore, these papers do not approach the ferry selection problem from an online perspective. Conversely, in this paper we propose an approach where each vehicle transfers a data bundle from source to destination without having to use relays and decisions are made in an online fashion---these assumptions are practical as more vehicles utilize OBU and GPS units that provide exact or probabilistic information about the path of the vehicle. Additionally, this paper considers online hiring algorithms for data ferry selection.

\section{System Model}

\begin{figure}[h]
\centering
\includegraphics[width=0.4\textwidth,trim={4 30 20 10},clip]{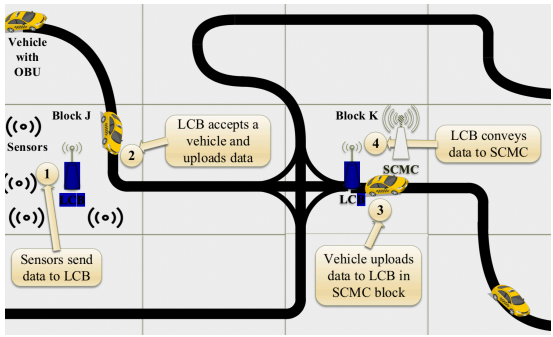}
\caption{Ferrying data from different blocks to the SCMC block in a community using vehicles.}
\label{fig:SystemModel}
\end{figure}

To utilize data ferrying in a given area, we divide the service area (e.g., city) into $B$ blocks. Each block has a number of smart devices (e.g., sensor nodes) that generate data. In addition, one of these blocks hosts the SCMC as shown in \figurename~\ref{fig:SystemModel}. Vehicles are used to transfer data collected from the smart devices to the SCMC. Also, each block has one LCB positioned near the center of the block. Any vehicle that enters a given block is within the communications range of its LCB.

In this paper, we make two assumptions about incoming vehicles. First, it is known if the vehicle will pass through the SCMC in the future. Second, for vehicles that pass through the SCMC, the expected arrival time is known. Our assumptions are based on many research efforts that appeared in the recent literature 
\cite{assumption_1, assumption_2, assumption_5, assumption_10} to predict those parameters.

Each LCB serves as a block manager that contacts incoming vehicles to acquire two pieces of information; namely, whether the vehicle is going to pass through the SCMC block, and at what time (i.e., expected arrival time at the SCMC block). If a vehicle is going to visit the SCMC block, the block manager computes its expected arrival time. Using the expected arrival time, the block manager computes the delivery delay $d$, which is the time required by the vehicle to transfer the data from the current block to the SCMC block. Besides, the block manager computes the waiting delay $w$, which is the time between the selection of the last vehicle and the selection of the current vehicle. In other words, $w$ is the time the data must wait until a vehicle is selected to serve as a data ferry.

When a vehicle passes through a block, the block manager, running the proposed algorithm, makes a decision on whether to accept or reject the current vehicle to serve as a data ferry. The block manager computes the delivery delay $d$ and the waiting delay $w$ and passes them to the proposed algorithm. Once the proposed algorithm makes a decision, it is impossible for that decision to be altered. The proposed algorithm utilizes an ensemble of online hiring algorithm as explained in Section \ref{heuristicSection}. If more than one vehicle exists in a block, and these vehicles are going to pass through the SCMC block some time in the future, the block manager considers only the vehicle with the minimum $d$.

Once a vehicle is selected, the LCB uploads data of block $j$ to the vehicle. Next, the vehicle continues its trip and eventually  passing through block $k$, which has the SCMC. Once in block $k$, the OBU of the vehicle uploads data collected from block $j$ to the LCB of block $k$, which in turn conveys it to the SCMC. These steps are illustrated in Fig~\ref{fig:SystemModel}.

\vspace{-5mm}

\section{Online Hiring Algorithms}

 Many companies around the world use hiring algorithms to select employees instead of the traditional manual selection process. Actually, there are many flavors of the hiring algorithm. In \cite{hiringPaper_2010} \cite{hiringPresentation_2007}, researchers investigate the performance of the different heuristics for the \textit{classical secretary problem} that select the best candidate out of multiple candidates. The problem involves an interviewer interviewing $n$ candidates one at a time for a position and then deciding after each interview if the interviewee is the best candidate. The overall goal in this problem that seeks to decide under uncertainty is to maximize the probability of choosing the best candidate. To select the best candidate, the authors introduce three hiring algorithms \cite{hiringPaper_2010} including (1) hire above a threshold, (2) hire above minimum or maximum, and (3) hire above mean or median (Lake Wobegon\footnote{Lake Wobegon refers to a fictional town, where ``\textit{all the women are strong, all the men are good looking, and all the children are above average}.''. A Lake Wobegon strategy refers to one that hires above the average (mean or median).}). We provide a brief description of these algorithms next.

\begin{itemize}

\item{\textit{Hiring Above a Threshold}}: In this version of the hiring algorithm, vehicle $i$ is selected only if the delivery delay $d_i$ is less than or equal to a fixed threshold $\tau$.

\item{\textit{Hiring Above the Mean}}: In this hiring algorithm version, vehicle $i$ is selected only if the delivery delay $d_i$ is less than $a_b$ (the average delivery delay of selected vehicles in block $b$). Initially, the algorithm accepts the first vehicle that enters block $b$ and then sets $a_b$ to $d_0$, and subsequently $a_b$ decreases gradually as the algorithm accepts more vehicles.

\item{\textit{Hiring Above the Median}}: Hiring above the median uses $m_b$ (the median of the delivery delay of all selected vehicles in block $b$). Like hiring above the mean, this algorithm initially accepts the first vehicle that enters block $b$. This algorithm needs an odd number of selected vehicles before recomputing $m_b$ since $m_b$ is the value in the middle after sorting. Therefore, after selecting a vehicle, if the number of selected vehicles is even, the algorithm does not update $m_b$ postponing the update of $m_b$ to the situation where the number of vehicles is odd.

\end{itemize}

\section{Heuristic Solution}
\label{heuristicSection}

In this work, we propose an algorithm that strives to minimize the average overall delay for transporting data from one block to the SCMC block. The overall delay is the sum of the waiting delay and the delivery delay. The idea is simply to run an ensemble of $N$ online hiring algorithms in passive mode while selecting only one of them to be active at any point in time. By passive, we mean an algorithm makes a decision for whether a given vehicle should be selected to serve as a data ferry but the decision is not executed. This is done in order to collect performance metrics needed to compare the performance of the different algorithms in the ensemble.

The proposed algorithm utilizes four hiring algorithms; namely, low threshold, high threshold, mean, and median. These algorithms can only consider the delivery delay and cannot take the waiting delay into consideration. In other words, they cannot make a decision based on the overall delay which includes the waiting delay. This stems from the fact that the waiting delay can be larger than the threshold used by those algorithms. Consequently, those algorithms will reject all requests after that time and will be stuck in this state forever. For example, if the threshold of the low threshold algorithm is set to 50 minutes and the algorithm is waiting for more than 50 minutes (i.e., waiting delay is greater than 50) then the overall delay will always be greater than 50 even if the delivery delay is zero. Thus, the low threshold algorithm will reject vehicles from serving as data ferries indefinitely. However, the proposed algorithm is capable of analyzing the history of all algorithms in its ensemble in terms of the overall delay. Moreover, to be efficient, the proposed algorithm has the chance to switch between the four algorithms in its ensemble every $S$ time units in case the performance of the already selected algorithm deteriorates.

Algorithm (\ref{alg_proposed}) shows the three parts of the proposed algorithm. One of the four algorithms in the ensemble is selected randomly to serve as the active algorithm in the initialization part, which is executed once when the algorithm starts. The second part is executed whenever a vehicle arrives and a decision needs to be made on whether to select it as a data ferry. In the second part, the four algorithms in the ensemble are executed and the overall delay of each one is saved for later analysis in the third part. Moreover, the decision made by the active algorithm is committed while decisions of other algorithms are ignored. The last part is only executed after $S$ time units had passed. Furthermore, the average overall delay of all algorithms is computed based on saved data and the algorithm with the minimum average overall delay is set as the active algorithm while setting the other algorithms as passive.

\begin{algorithm}[!t]
\footnotesize
\caption{Proposed algorithm for selecting ferries }
\label{alg_proposed}
\begin{algorithmic}[1]
\STATEx \textbf{Input}: vehicle arrival time $A$.
\STATEx \textbf{Output}: decision (accept or reject)

\STATEx  \textit{\textbf{Initialization} (executed once at time 0)}:
\STATE Set all algorithms as passive
\STATE Set activeAlgorithm = select an algorithm randomly.

\STATEx  \textit{\textbf{On vehicle arrival}:}
\STATE Set $d=A-$ current time
\STATE Set $w=$ current time - last time a vehicle was accepted
\FOR {each of the 4 algorithms}
\STATE Run the algorithm
\IF {Decision is accept}
\STATE Save overall delay as $d+w$
\IF {this algorithm is activeAlgorithm}
\STATE Accept the vehicle
\ENDIF
\ELSIF {this algorithm is activeAlgorithm}
\STATE Reject the vehicle
\ENDIF
\ENDFOR
\STATEx  \textit{\textbf{Executed every $S$ minutes}:}
\STATE Compute average overall delay based on saved data
\STATE Set bestAlgorithm = algorithm with minimum delay
\IF {activeAlgorithm $\neq$ bestAlgorithm}
\STATE Set activeAlgorithm = bestAlgorithm
\ENDIF
\end{algorithmic}
\end{algorithm}

\section{Illustrative Example}

Let Algorithm A and Algorithm B be two online hiring algorithms with Algorithm A initialized in passive mode and Algorithm B initialized in active mode (i.e., selected randomly by the proposed algorithm) at $t_0$. Table~\ref{table_example} shows the average overall delay per algorithm recorded every $S$ minutes (see Algorithm \ref{alg_proposed}), which is computed as the average of overall delays in the period $t_i$ to $t_j$, where $j=i+1$. The performance of the proposed algorithm is not the best at $t_1$ since it randomly selects Algorithm B as the active algorithm between $t_0$ and $t_1$, which performs worst than Algorithm A. However, the proposed algorithm switches to Algorithm A at $t_1$ and got an average overall delay of 4 minutes between $t_1$ and $t_2$, which is the same value gained by Algorithm A. Between $t_2$ and $t_3$, Algorithm B performs better than Algorithm A leading the proposed algorithm to get 6 minutes instead of 2 minutes. In $t_3$, the proposed algorithm switches to Algorithm B and get the minimum average overall delay of 2 minutes between $t_3$ and $t_4$. By $t_4$, the proposed algorithm achieves less average overall delay compared to both algorithms.

\begin{table}
\footnotesize
	\centering
	\caption{Performance (average overall delay) of the proposed algorithm compared to the two online hiring algorithms}
	\label{table_example}
    \begin{tabular}{|c|c|c|c|c|}
        \hline
        \multirow{2}{2em}{\thead{Time}} &
        Algorithm A &
        Algorithm B &
        \multicolumn{2}{c|}{Proposed} \\
        \cline{2-5}
        & \thead{Avg. delay} & \thead{Avg. delay} & \thead{Avg. delay} & Selection \\
        \hline
        $t_0$ & 0 & 0 & 0 & Algorithm B \\
        \hline
        $t_1$ & 8 & 10 & 10 & Algorithm A \\
        \hline
        $t_2$ & 12 & 22 & 14 & Algorithm A \\
        \hline
        $t_3$ & 18 & 26 & 20 & Algorithm B \\
        \hline
        $t_4$ & 30 & 28 & 22 & Algorithm B \\
        \hline
    \end{tabular}
\end{table}

\section{Experimental Results}

In this section, we describe the dataset used in our experiments; explain the experiments' settings; evaluate the performance of the proposed online algorithm by comparing it with four baseline online hiring algorithms using real vehicular traces; and finally, discuss the results and present the major insights learned from our experiments.

\subsection{Dataset \& Experimental Settings}

In our experiments, we make use of the Shanghai dataset consists of taxi traces collected in the city of Shanghai in China. Each taxi has a GPS unit and a GPRS wireless communications modem. Vehicles send their GPS location along with other information to a data center every minute. Around 2,109 taxis participated in this dataset in 2007. Information sent by taxis includes ID, timestamp, longitude and latitude, speed, and heading direction \cite{shanghai_2007}.

In order to utilize the Shanghai dataset, we have encoded the geographical location that encompasses longitude and latitude into a string of 7 characters using the \texttt{GeoHashing} method~\cite{geoHashing_2015}. 
Every string represents a grid (i.e., block) of the city. Actually, we used a \texttt{GeoHashing} of 7 characters because this allows for the division of the globe into blocks, each of $153 \times 153$ meters, which is within the communications coverage of a typical LCB. Using these blocks, we position the SCMC in the block that is mostly visited by vehicles. Additionally, we filtered the dataset to remove blocks that have no traffic activity and focus on the active blocks. The dataset is based on a one-day observation. However, one day is a very short period for the proposed algorithm to work effectively. Therefore, we replicate the one-day data a number of times to have datasets for 5, 10, 15, 20, and 25 continuous days. 

To study the performance of the proposed algorithm under different traffic scenarios, we divide the city into three areas based on the traffic volume. We computed the average number of vehicles per block along with the standard deviation and found that the standard deviation is greater than the average. Therefore, we categorized each block based on $N_b$, the number of vehicles in block $b$, as follows:
\begin{itemize}
    \item \textit{Light traffic area}: $N_b <$ average
    \item \textit{Medium traffic are}a: average $\leq N_b \leq$ standard deviation 
    \item \textit{High traffic area}: $N_b >$ standard deviation
\end{itemize}

We set $S$ to 30 minutes in all of the experiments to be consistent. Also, to derive the low and high threshold values for the threshold algorithms in every block, we used a percentile of the delivery delay in the block. The dataset used in the experiment has a heavy-tailed distribution and particularly a long-tail distribution and we resort to extremely low and high threshold values---\nth{2} percentile for the low threshold algorithm and \nth{95} percentile for the high threshold algorithm---to fully explore the space of values in such a distribution.



\subsection{Results Discussion}

\subsubsection{Evaluation of the proposed algorithm for different traffic volumes}

\begin{figure}[!th]
\centering
\includegraphics[width=.40\textwidth]{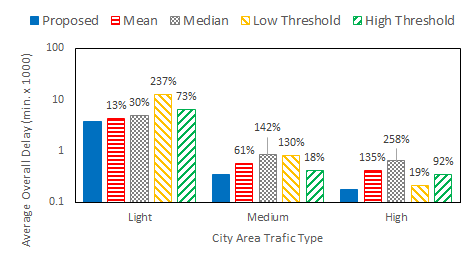}
\caption{Average overall delay per block. \textit{Our proposed algorithm achieves minimal overall delay per block regardless of the traffic volume.}}
\label{fig:10Days_overall}
\end{figure}

Considering the four baseline hiring algorithms only, it can be clearly seen that a different one outperforms in each area depending on the traffic volume. The mean algorithm is the best in light traffic areas, the high threshold is the best in medium traffic areas, and the low threshold is the best in high traffic areas as shown in Fig.~\ref{fig:10Days_overall}.

\begin{figure}[!th]
\centering
\begin{subfigure}[b]{0.40\textwidth}
\includegraphics[width=\textwidth]{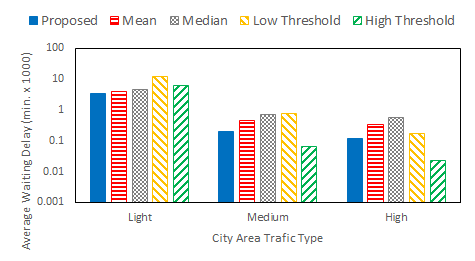}
\caption{Average waiting delay per block.}
\label{fig:10Days_waiting}
\end{subfigure}
\begin{subfigure}[b]{0.40\textwidth}
\includegraphics[width=\textwidth]{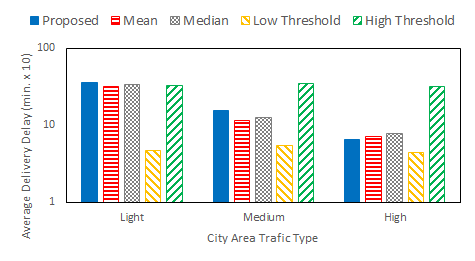}
\caption{Average delivery delay per block.}
\label{fig:10Days_delivery}
\end{subfigure}
\caption{Average waiting \& delivery delay per block. \textit{Our proposed algorithm performs the best either with minimal average waiting delay or average delivery delay per block, but not with both.}}
\label{fig:Delays}
\end{figure}

\begin{figure}[!th]
\centering
\includegraphics[width=.40\textwidth]{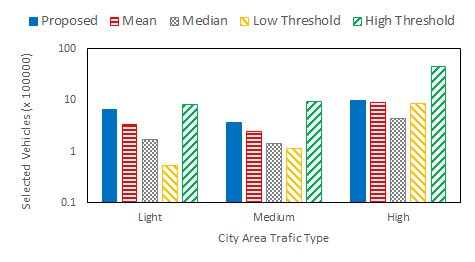}
\caption{Number of selected vehicles. \textit{Our proposed algorithm hires the most number of vehicles with the exception of the high threshold algorithm.}}
\label{fig:10Days_selected}
\end{figure}

\begin{figure*}[htbp]
\centering
\begin{subfigure}[b]{0.32\textwidth}
\includegraphics[width=\textwidth]{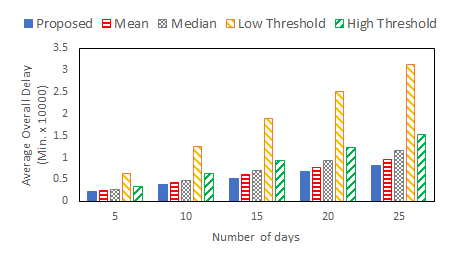}
\caption{Light traffic area.}
\label{fig:allDays_light}
\end{subfigure}
\begin{subfigure}[b]{0.32\textwidth}
\includegraphics[width=\textwidth]{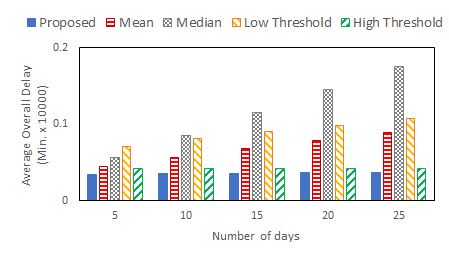}
\caption{Medium traffic area.}
\label{fig:allDays_medium}
\end{subfigure}
\begin{subfigure}[b]{0.32\textwidth}
\includegraphics[width=\textwidth]{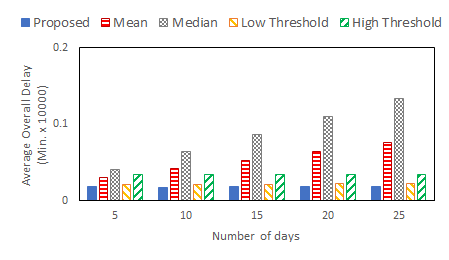}
\caption{High traffic area.}
\label{fig:allDays_high}
\end{subfigure}
\caption{Average overall delay using different algorithms for the various traffic scenarios (light, medium, high) over different days. \textit{Our proposed algorithm achieves the minimal overall delay for different number of days results regardless of the traffic volume.}}
\label{fig:allDays}
\vspace{-5mm}
\end{figure*}

\begin{figure}[htbp]
\centering
\includegraphics[width=0.43\textwidth,trim={0 0 0 -5},clip]{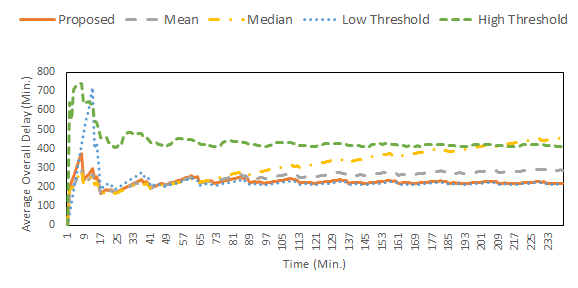}
\caption{Performance of algorithms in one block over 10 days. \textit{Our proposed algorithm switches between different hiring algorithms to achieve minimal overall delay.}}
\label{fig:Switching}
\end{figure}

The low threshold algorithm suffers from a high overall delay in the light traffic areas because the waiting delay is very high since it is very selective. However, the low threshold algorithm outperforms in the area of the high traffic since it only selects vehicles with low delivery delay and there are plenty of vehicles to pick from. On the other hand, the high threshold algorithm performs best in the medium traffic areas, which provide a balance between waiting delay and delivery delay. Since the high threshold algorithm accepts the majority of vehicles, it benefits from this balance. As for the mean and median algorithms, their performance is best in the areas of light traffic. This is because the thresholds of these algorithms decrease with more vehicles. The more these algorithms accept, the more greedy they become towards a lower threshold.

The proposed algorithm achieves the best results in all areas with up to 258\% less overall delay albeit at a cost. To understand this cost, we focus on the 10 days results and record the number of selected vehicles, average delivery delay, average waiting delay, and average overall delay.

The proposed algorithm outperforms the baselines algorithms regardless of the traffic volume by either performing better on the waiting delay or on the delivery delay but not both as indicated in Fig.~\ref{fig:Delays}.

It should be noted that the proposed algorithm does not only perform better in terms of the average overall delay but it also accepts more vehicles to serve as data ferries as shown in Fig.~\ref{fig:10Days_selected} (with the exception of the high threshold algorithm since it accepts the majority of vehicles in all areas).

\subsubsection{Evaluation of the proposed algorithm for different time periods}

To assess the performance of proposed algorithm relative to the four baseline hiring algorithms, we run the algorithms for different number of days. Results are collected in terms of the average overall delay in each of the three different areas as shown in Fig.~\ref{fig:allDays}. The figure shows the consistent behaviour of the proposed algorithm regardless of the number of days.

\subsubsection{Switching activity of the proposed algorithm}
To show the switching activity of the proposed algorithm, we record the average overall delay every hour for one block over 10 days as illustrated in Fig.~\ref{fig:Switching}. The figure shows how some algorithms perform better for a period of time and how the proposed algorithm follows the one with the minimum average overall delay based on performance collected from the recent history.

\section{Conclusions and Future Work}

In this paper, the problem of selecting vehicles to serve as data ferries in support of smart community applications is considered. The selection process strives to achieve the minimum average overall delay. An online algorithm is proposed that utilizes four online hiring algorithms by running all of them together in passive mode and selecting the one that has performed the best in recent history. The proposed algorithm is evaluated using real taxi traces from the city of Shanghai in China and compared against a baseline of four online hiring algorithms. Experiments with these traces demonstrate that the proposed algorithm outperforms online hiring algorithms presented in the literature regardless of the traffic volume by either performing better on the waiting delay or on the delivery delay but not both.

In the future, we plan to evaluate the proposed algorithm analytically to provide performance guarantees, in terms of competitive ratio, in worst-case scenarios.

\ifCLASSOPTIONcaptionsoff
  \newpage
\fi

\bibliographystyle{IEEEtran}
\bibliography{biblo}

\end{document}